\documentclass{iopart}
\usepackage{epsfig,graphicx}

\begin{document}

\maketitle

\title{Lattice effects and current reversal in superconducting ratchets}

\author{L. Dinis$^{{1}}$ , E.M. Gonz\'{a}lez$^{{2}}$ , J. V. Anguita$^{{3}}$
, J.M.R. Parrondo$^{{1}}$ , J.L. Vicent$^{{2}}$ }

\address{$^{{1}}$ Grupo Interdisciplinar de
Sistemas Complejos (GISC) and Departamento de F\'isica At\'omica,
Nuclear y Molecular. Universidad
Complutense de Madrid. E-28040 Madrid, Spain.\\
$^{{2}}$ Departamento de F\'isica de Materiales. Universidad
Complutense de Madrid. E-28040 Madrid, Spain.\\
$^{{3}}$ Instituto de Microelectr\'onica de Madrid. Consejo Superior
de Investigaciones Cient\'ificas. Tres Cantos, E-28760,Spain.}

\begin{abstract}
Competition between the vortex lattice and a lattice of asymmetric
artificial defects is shown to play a crucial role in ratchet
experiments in superconducting films. We present a novel and
collective mechanism for current reversal based on a reconfiguration
of the vortex lattice. In contrast to previous models of vortex current
reversal, the mechanism is based on the global response of the
vortex lattice to external forces.
\end{abstract}


\section{Introduction}

The term {\em ratchet effect} refers to a net flow of
 particles induced by asymmetric potentials, in the
absence of non-zero average forces. This effect has drawn the
attention of many researchers from very different fields
\cite{peter}. It is believed to play a role in protein motors
\cite{prostcollective} and it has been used to design synthetic
molecular motors \cite{leigh,aron} and control the motion of
colloidal particles \cite{separation} .

In the context of superconductivity,  vortices in superconducting
films \cite{barabasi,nori} and Josephson junctions
\cite{zapata,marconi} were soon proposed as implementations of the
ratchet effect. The first experimental evidence of rectification of
vortices was reported by Villegas et al. \cite{science}. In this
work, a Nb superconducting film was grown on an array of triangular
Ni nanoislands, acting as pinning sites for the vortices. When the
vortex lattice is driven by an input ac-current, the asymmetric
geometry of the pinning potential leads to a net flow of vortices,
resulting in an output dc-voltage.  After this seminal work, considerable
theoretical \cite{Nori1,Nori2,Nori3} and experimental \cite{PRLMoshchalkov} research
has been done in order to clarify ratchet mechanisms or use them to control the motion of vortices in films and Josephson arrays.

An interesting feature of this experimental setup is the appearance
of current reversals, i.e., the rectification polarity changes for
certain values of the applied magnetic field and current strength.
This behavior has been explained in Ref. \cite{science} by means of
a one dimensional model, where current reversal is due to the
interplay between vortices pinned in the asymmetric centers
(triangles) and interstitial vortices placed among the triangles.
However, in that model the interaction between pinned and interstitial
vortices was replaced by an effective ratchet potential affecting only to
interstitial vortices. Beside this effective potential, both pinned and
interstitial vortices were treated as independent particles.

In the present paper, we review the experimental results of Ref.~\cite{science} of rectification and current reversal using magnetic defects, and present
experimental results showing that the current reversal persists with
non-magnetic defects. Furthermore, we introduce a new theoretical approach
to the current reversal considering interaction between vortices. Furthermore, we introduce a new theoretical approach
to the current reversal considering interaction between vortices.
In the absence of pinning potentials, it is well
known that the ground state induced by vortex interaction is a
triangular Abrikosov lattice \cite{tinkham}. Our aim is to analyze
how this vortex lattice is distorted by the lattice of asymmetric
defects. The competition between the two lattices will be relevant
when the magnetic field creates a few vortices per pinning site and
for long penetration depth, $\lambda $, which is the case for
temperatures close to the critical temperature, as in the
experiments reported in \cite{science}. Furthermore, we will show
that this vortex lattice distortion can induce current reversal but
nevertheless the lattice exhibits a net motion, i.e., both pinned
and interstitial vortices move in the same direction. Therefore, our
picture of vortex rectification and current reversal differs both from the one presented in Ref. \cite{science}, where the latter is induced only by the
motion of interstitial vortices, and that in Refs. \cite{Noriprl9192,Nori1}
 which deal with systems of particles where adding particles of an auxiliary species, that interact with both the primary particles of interest and the substrates, the particles could move either together or in opposite directions depending upon the strength of the interaction, and whether it is attractive or repulsive.

Moreover, in our new approach, the superconductor ratchet becomes a
collective ratchet. Collective effects in ratchets have been shown
to yield new interesting phenomena such as instabilities and
hysteresis \cite{prostcollective}, absolute negative resistance
\cite{anr} and inefficiency of some feedback control protocols
\cite{control}. Moreover, many biological motors, such as kinesins
and myosins work in a collective way \cite{prostcollective}.
Collective effects have been also considered in the context of
superconducting vortices by Reichhardt et al.~\cite{PhysicaC} and de
Souza Silva et al.~\cite{nature}. Also Shalom and Pastoriza suggested that collective effects are responsible for current reversal in Josephson junction arrays \cite{Shalom}. 

In our approach we show that the ratchet reversal corresponds to a reconfiguration of the vortex lattice at different fields and drives. Similar arguments about the importance of the vortex-vortex interactions in current reversal mechanisms were made in Ref.~\cite{PhysicaC}. In their simulations they observe how interstitial vortices may help depin vortices at the tip of the triangular defect due to thermal fluctuations but not the ones at the base of the triangle, leading to a negative current at low drives.  Also, the current reversal is explained by vortex-vortex interactions in the framework of a one-dimensional model in \cite{nature}.
 However, none of these works discuss the
aforementioned competition between the Abrikosov lattice and the
array of defects.

Recently, Lu et al.~\cite{Lu} have reported a vortex ratchet effect for a 2D
vortex system interacting with a substrate that has an asymmetric
periodic modulation along one direction. Although the pinning in that
system might be expected to have a 1D character, Lu et al showed that
the ratchet effect and ratchet reversals that arise are produced by
collective vortex-vortex interactions, and also showed that the
overall vortex lattice structure can change for drives in the opposite
direction.

The paper is organized as follows. In section 2 we review the
experiment and present results with Cu non-magnetic artificial
defects. In section 3 we introduce the model used in numerical
simulations. In section 4 we discuss current reversal and lattice
configurations. Finally, we present our conclusions in section 5.

\section{Experimental method}
%

We have fabricated the Nb thin films on top of arrays of Cu or Ni
nanotriangles grown on Si (100) substrates, by means of electron
beam lithography, magnetron sputtering and ion etching techniques
\cite{science,PRBVillegas}. The array aspect ratio and the triangle
dimensions of the samples are similar to the films reported by
Villegas et al. \cite{science}: the periodicity of the array is
$770\,{\rm nm}\times 746\,{\rm nm}$, the triangle base is $620\,{\rm
nm}$ and height is $477\,{\rm nm}$
. A cross-shaped patterned
bridge allows us to inject a current into the sample and to measure
the voltage drop parallel or perpendicular to the triangle base. In
all the experiments, the magnetic field $H$ is applied perpendicular
to the substrate. 
In the ratchet measurements an injected ac current
is applied and the output dc voltage is recorded, whereas a
commercial He cryostat is used to control the temperature. More
fabrication and characterization details have been reported elsewhere \cite%
{science,PRBVillegas}.

The dc magnetoresistance of superconducting thin films with periodic
arrays of pinning centers show minima when the vortex lattice
matches the unit cell of the array \cite{PRL97Martin}. These minima
are sharp (strong reduction of the dissipation) and equally spaced
(two neighbor minima are always separated by the same magnetic field
value). Consequently, the number of vortices per array unit cell can
be known by simple inspection of the dc magnetoresistance $R(H)$
curves, in which the first minimum corresponds to one vortex per
unit cell, the second minimum to two vortices per unit cell, and so
on. Moreover, the ratio between the dimension of the pinning center
and the superconducting coherence length governs the maximum number
of vortices that could be pinned in each one of the pinning centers
\cite{Shmidt}. The dimensions of our defects allow a maximum of
three pinned vortices per triangle. Therefore, increasing the
applied magnetic field to four vortices per array unit cell leads to
three pinned vortices and one interstitial vortex \cite{science}. In
general, we know, for selected values of the applied magnetic field,
how many vortices are per unit cell and how many of them are pinned
and interstitial in the ground state and for zero external force. We
have to notice that, in this type of experiments, the periodic
potentials govern the vortex dynamics inducing many remarkable
effects, as vortex lattice reconfiguration \cite{PRL99Martin} or
vortex lattice channeling effects \cite{VelezPRB02}. 
Finally, we have used different frequencies, up to 10kHz, for the
input signal with identical results. This indicates that our
experiments are done in the adiabatic regime
\cite{PRBVillegas,PRLMoshchalkov}, i.e., the period of the signal is
much longer than the relaxation time of the system, which is able to
adopt the corresponding stationary state at any time.

\begin{figure}[tbp]
\begin{center}
\includegraphics[width=7cm]{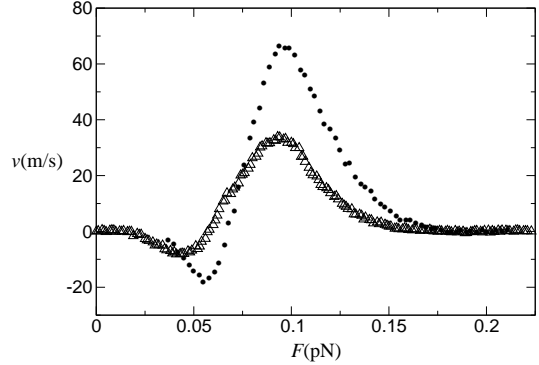}
\end{center}
\caption{Velocity vs. driving force for an  applied magnetic field
is $H= 1.3\times 10^{-2}$ T ($n=4$ vortices per array unit cell) and
temperature $T/T_{c}= 0.99$. The ac current frequency is 10 kHz. The
sample is a Nb film $100\,{\rm nm}$ thick with array of $40\,{\rm
nm}$ thick triangles (array periodicity $770\,{\rm nm}\times
746\,{\rm nm}$, triangle side $620\,{\rm nm}$). Dots: Nb film with
array of Ni triangles. Triangles: Nb film with array of Cu
triangles.} \label{fig:experimento}
\end{figure}

Figure \ref{fig:experimento} shows the experimental results of Nb
films on arrays of Ni (magnetic)  and Cu (non-magnetic)
nanotriangles for $n=4$ vortices per triangle. The results
corresponding to Ni defects are similar to those reported in Ref.
\cite{science} with a clear current reversal. The Cu non-magnetic
defects exhibit a similar behavior. In this type of defect, pinning
is weaker \cite{hoffmannPRB} and, consequently, rectification is
lower and depinning forces are smaller.

\section{Theoretical model}

To clarify the mechanisms behind the experimental results, we have
performed extensive numerical simulations of vortices as a set of
interacting and overdamped two-dimensional Brownian particles.
Similar simulations have been used to study the DC vortex transport properties of  arrays of triangular pinning antidots \cite{noriPhysicaE} and current reversal in arrays of triangular pinning defects \cite{PhysicaC}, although in these references the effects of possible lattice reconfigurations are not analysed.

 The corresponding Langevin equation for position $\mathbf{r}_i$ of
vortex $i$ reads:
\begin{equation}
\eta \dot{\mathbf{r}_i}(t)=-\sum_{j\neq
i}\vec{\mathbf{\nabla}}_{i}U_{vv}\left(\left|{\bf r}_i-{\bf
r}_j\right|\right)+ \mathbf{F}_{\mathrm{pinn}}({\bf
r}_i)+\mathbf{F}_{\mathrm{ext}}(t)+\mathbf{\Gamma}_i (t)
\label{eq:langevin}
\end{equation}
where $\eta$ is the viscosity, $U_{vv}(r)$ is the interaction
potential between vortices, $\mathbf{F}_{\mathrm{pinn}}({\bf r})$
the force due to pinning artificial defects,
$\mathbf{F}_{\mathrm{ext}}(t)$ the external force due to the applied
currents, and  $\mathbf{\Gamma}_i(t)$ are white Gaussian noises
accounting for thermal fluctuations: \begin{equation}
 \langle \mathbf{\Gamma}_i(t)\cdot\mathbf{\Gamma}_j(t^{\prime })\rangle ={4kT}{\eta}\delta_{ij}
 \delta (t-t^{\prime }).
\end{equation}
For the vortex interaction potential, we have used \cite{tinkham}:
\begin{equation}
U_{vv}(r)=\frac{\phi _{0}^{2}}{4\pi \mu _{0}\lambda ^{2}}K_{0}\left( \frac{%
r}{\lambda }\right)  \label{eq:vv}
\end{equation}%
where $\phi _{0}=hc/(2e)$ is the quantum of flux, $\mu_0$ is the
magnetic permeability of vacuum, $K_0$ is the zeroth-order modified
Bessel function, and $\lambda$ is the penetration depth of the
material, which is very sensitive to temperature for the conditions
of the experiment, very close to the critical temperature.

The external force $\mathbf{F}_{\rm{ext}}$ is the Lorentz force
resulting from the applied current. As in the experiment, this force
is perpendicular to the basis of the triangles and sinusoidal in
time.

\begin{figure}[tbp]
\center
\includegraphics[width=8cm]{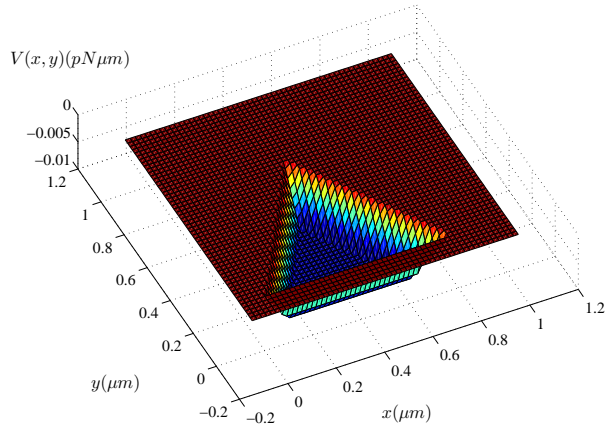}
\caption{Potential acting on vortices due to pinning defects.}
\label{fig:potencial}
\end{figure}

The triangular pinning defects have been simulated using an
attractive potential in the shape of an inverted triangular
truncated pyramid, that is, null force in the middle section of the
triangle, and constant force $F_{p}$ pointing inward in a region of
width $d$ along the walls of the pinning defect (see figure
\ref{fig:potencial}). We have used an array of 6$\times 6$ defects
with periodic boundary conditions.

The parameters in the simulation  have been chosen similar to
already reported values in samples of Nb films with arrays of
periodic pinning centers \cite{MartinPRB00}. We take $\lambda
=360\,{\rm nm}$ and temperature $T=8.26\,{\rm K}$, which are typical
values close to the superconducting critical temperature. The
pinning strength
$F_{p}=0.12\,{\rm pN}$ and the width of the pinning potential $%
d=0.7\,\mu \mbox{m}$ have been set so that vortices start depinning
approximately at the same applied force as observed in the
experiment (see \cite{science}), whereas the dimensions of the
triangles in the simulation correspond to those in the experiment.
We point out that nevertheless simulations performed with small 
variations on these parameters exhibit similar behavior.

Finally, we have followed the suggestion made in \cite{velez}
introducing a viscosity $\eta(F)$ depending on the strength of the
external force. The reason of this dependence is that viscosity is
in fact the result of the interaction of vortices with random
intrinsic pinning centers scattered all over the material.  To mimic
the behavior of the viscosity measured in Ref. \cite{velez}, we have
taken in our simulations the following phenomenological expression
for the viscosity:
\begin{equation} \eta(F) =\left(
0.005\,e^{-100|F({\rm pN})|}+6\times 10^{-6}\right) \mbox{
pN$\cdot$s/m}.
\end{equation}
Notice however that the viscosity enters in the Langevin equation
(\ref{eq:langevin}) as a time scale. Therefore, it affects the
magnitude of the velocity of the vortices, but not the qualitative
behavior of the lattice or the mechanisms that lead to current reversal. 

\begin{figure}[tbp]
\center
\includegraphics[width=7cm]{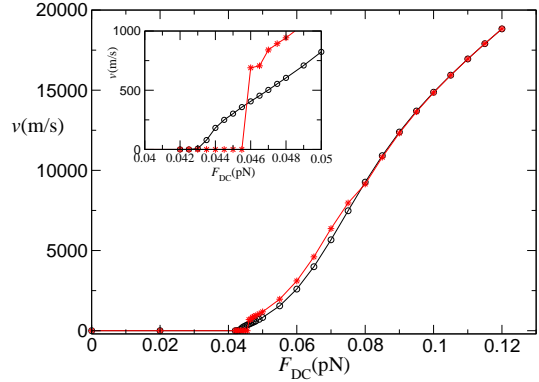}
\caption{Velocity (absolute value) of the lattice as a function of the
applied force, for positive (*) and negative (o) forces. Simulation results
and linear interpolation. Inset shows the inversion in more detail.}
\label{fig:DC}
\end{figure}

 As mentioned  above, our experimental results, in agreement with previous work
 \cite{PRBVillegas,PRLMoshchalkov}, show that the system is adiabatic in the
region of applied frequencies, which allowed us to perform our
simulations also in the adiabatic limit. We first obtain $v_{dc}(F)$
curves for different values of constant positive and negative forces
(dc curves). These dc curves are depicted in figure \ref{fig:DC},
where the absolute value of the velocity of the vortex lattice is
plotted against the applied force. The inversion is clearly shown in
the inset. Once the dc curves are obtained, the corresponding ac
curves $v_{ac}(F_{ac})$ can be easily computed by performing the
following numerical integration which does not depend on the
frequency of the force (along the paper $F_{ac}$ stands for the
r.m.s. value of the external force):
\begin{equation} v_{ac}(F_{ac})=\int_0^{2\pi}v_{dc}\left(F_{ac}\sqrt{2}\sin
(t)\right)dt.
\end{equation}

\begin{figure}[tbp]
\center
\includegraphics[width=7cm]{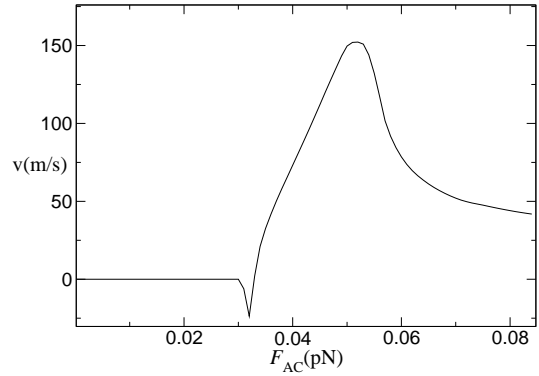}
\caption{Average velocity of the vortices as a function of the
r.m.s. amplitude $F_{ac}$ of the external force.}
\label{fig:n4adiab}
\end{figure}

Figure \ref{fig:n4adiab} shows the average velocity for 144
particles and $6\times 6$ pinning triangles ($n=4$) and the
parameters described above, as a function of the external force
$F_{ac}$. Simulations reproduce the experimentally observed current
reversal, i.e., negative current for small forces and positive for
higher forces, decaying to zero afterward. Furthermore, the forces
applied and the corresponding velocities are both quite similar to
those in the experiment.

\section{Discussion: vortex lattice reconfiguration and current reversal}

Consider a single particle or vortex in the triangular potential of Fig. \ref%
{fig:potencial}. The force to depin the vortex when pushing  in the
positive $y$ direction through the tip of the triangle is
approximately half the force needed to depin it through the base,
when pushing in the negative $y$ direction. Positive or upward is
therefore the ``natural'' direction for the rectification induced by
triangular defects, and it is observed experimentally for $n=1,2,3$
\cite{science}. However, as shown in Fig. \ref{fig:experimento}, for
$n=4$  downward rectification is also observed for small forces, and
the same unexpected result is obtained for $n\ge 4$ \cite{science}.

What is the mechanism of this negative rectification? Previous
models \cite{science} consider that interstitial vortices feel an
effective ratchet potential which is roughly the spatial inverse of
the pinning potential, so they are rectified in the negative
direction and move independently of the pinned vortices. However, a
simple visual inspection of our simulations reveal that the lattice
moves as a whole without significant shear and that current reversal
is due to a reconfiguration of the vortex lattice (see supplementary
multimedia material).

In figure~\ref{fig:fundamental}(a), we depict  the vortex lattice
that minimizes the total free energy and has the periodicity of the
rectangular array of triangles, for $n=4$  and zero external force.
We have found this configuration by simulated annealing  on a cell
with a single defect and periodic boundary conditions. Notice that
the actual ground state or minimal energy configuration might not in
principle have the periodicity of the array of defects, specially
for weak pinning forces. However, our simulations reveal that this
is not the case for the parameters chosen above:
figure~\ref{fig:fundamental}(a) is indeed the ground state of the
system, consisting of a slightly distorted triangular lattice
(vortices form isosceles triangles but no equilateral) with one
interstitial and three pinned vortices.

This lattice starts to move either upward or downward if we add a
force, but it turns out that the depinning force is stronger in the
positive direction, yielding a negative rectification.

\begin{figure}[tbp]
\center
\includegraphics[width=\textwidth]{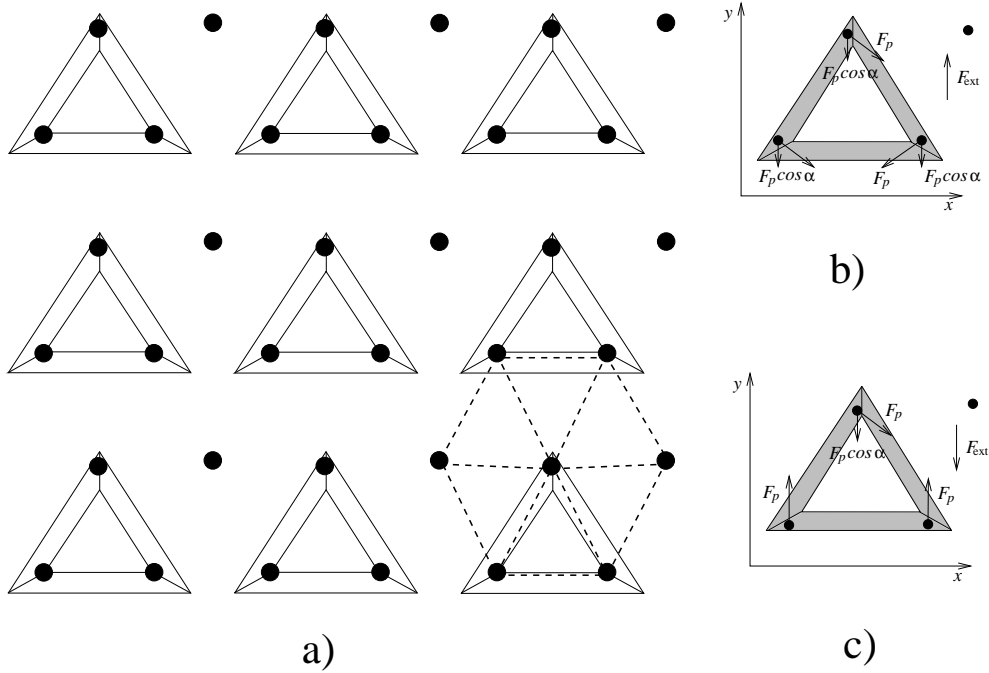}
\caption{The ground state of the system for zero external force.}
\label{fig:fundamental}
\end{figure}

The effect can be explained by considering first a rigid lattice.
When pushing upward (toward the positive $y$ direction) the $y$
component of the total force acting on the three pinned vortices due
to the pinning is $-3F_{p}\cos\alpha$ (Fig.
\ref{fig:fundamental}(b)), $\alpha$ being the angle at the basis of
the triangle. On the other hand, for a negative motion, the total
force to overcome in the $y$ direction is
 $(2-\cos\alpha)F_{p}$ (Fig.~\ref{fig:fundamental}(c)). In the case
 of  equilateral triangles, $\cos\alpha=1/2$
 and both depinning forces are identical. Our triangles are however
 a little bit flattened,
 with base $620\,{\rm nm}$ and height $477\,{\rm nm}$. Using this values
 and  $F_p=0.12\,{\rm pN}$, the resulting depinning forces per
vortex are $0.0437\,{\rm pN}$ (downward) and $0.0490\,{\rm pN}$
(upward).

Although this argument is only valid for zero temperature and a
rigid lattice, the depinning forces observed in the simulation (Fig.
\ref{fig:n4adiab}) are very close to these theoretical values: we
observe downward depinning for $0.0435\,{\rm pN}$ and upward for
$0.0460\,{\rm pN}$. The greater discrepancy for the upward motion
indicates that the plasticity of the lattice is more relevant in the
upward depinning, whereas the lattice is barely deformed in the
downward motion. In the supplementary multimedia material one can
see how the whole lattice remains pinned for an external force
$F=0.44\,{\rm pN}$, whereas it does move downward for the same
negative force $F=-0.44\,{\rm pN}$ with a small deformation. We then
conclude that the negative rectification is mainly due to the shape
of the triangular defects.

Increasing the external force we get upward rectification. For
 $F=0.5\,{\rm pN}$, the lattice is already depinned in the
upward direction, but we observe a very interesting phenomenon which
is the main result of the paper: the lattice undergoes a transition
(see supplementary
multimedia material) adopting the configuration depicted in figure~\ref%
{fig:fuerzapos} with only one (or two at most) pinned vortices. This
new configuration is the result of a 90$^{\rm o}$ rotation of the
ground state of Fig.~\ref{fig:fundamental}, and induces  a
relatively fast columnar motion of vortices which exceeds the motion
in the downward direction for the same magnitude of the external
force, yielding a net upward rectification.

\begin{figure}[tbp]
\center
\includegraphics[width=0.75\textwidth]{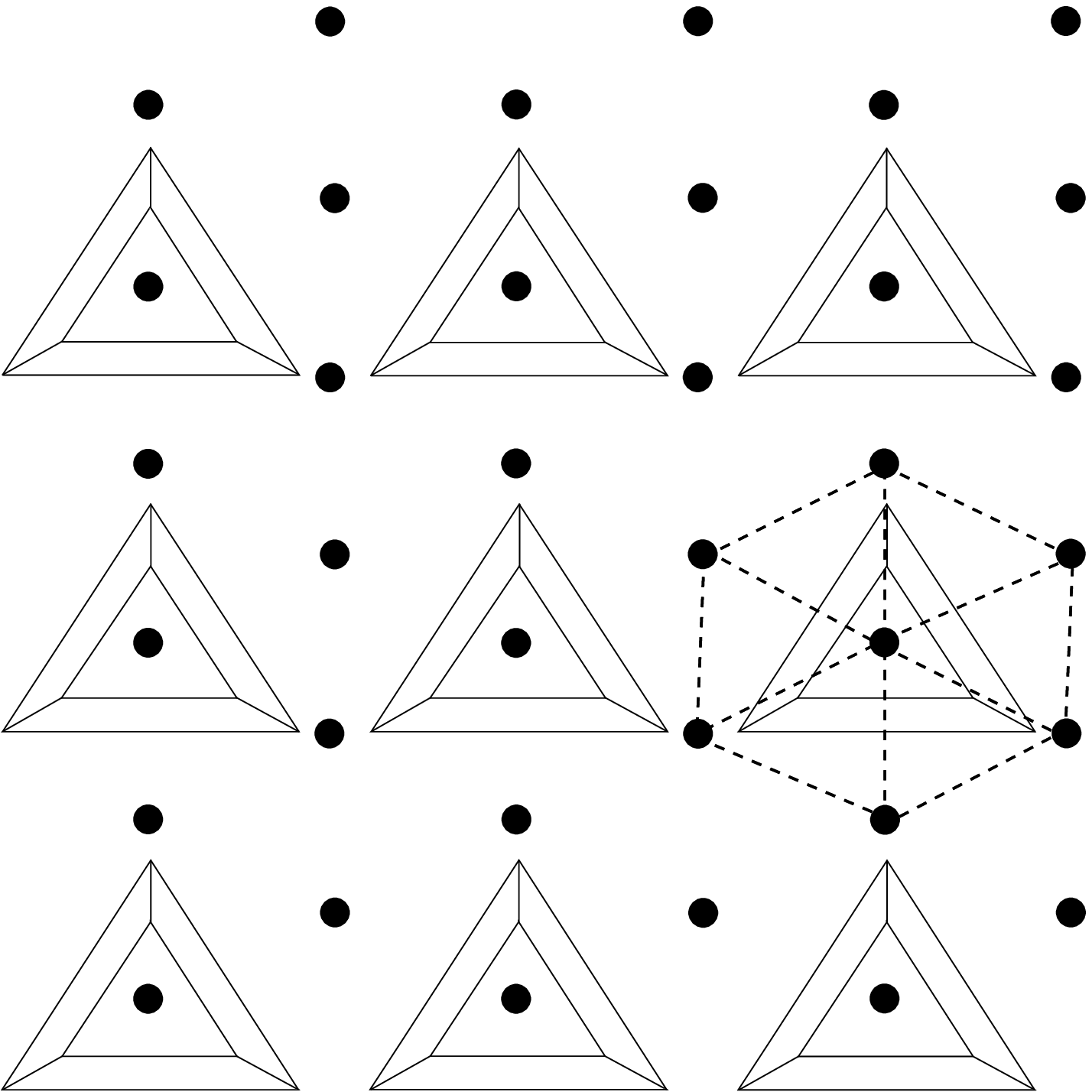}
\caption{Configuration for $F_{\mathrm{ext}}=0.1\,{\rm pN}$.}
\label{fig:fuerzapos}
\end{figure}

Therefore, our simulations reveal that the current reversal is
entirely due to a reconfiguration of the vortex lattice. Changes in
lattice configuration have been extensively reported in the
literature, mainly in equilibrium
\cite{PRL99Martin,reconfig1,reconfig2}. In our case, however, the
reconfiguration is a dynamical phenomenon. In spite of this, we can
gain some insight on the problem by calculating the interaction
energy of the configurations depicted in Figs.~\ref{fig:fundamental}
and \ref{fig:fuerzapos} and comparing it with the interaction energy
of triangular and square lattices with the same density of vortices.

Let $a$ and $b$ be  respectively the horizontal and vertical
distance between triangles in the array of defects. The lattice in
Fig.~\ref{fig:fundamental} (we will call it H lattice) is given by
$n{\bf v}_{H,1}+m{\bf v}_{H,2}$, with $n,m=0,\pm 1,\pm 2,\dots$ and
the following generating vectors:
\begin{equation}
{\bf v}_{H,1}=\left(\begin{array}{c} a/2 \\ 0
\end{array}\right) ;\quad {\bf v}_{H,2}=\left(\begin{array}{c} a/4 \\ b/2
\end{array}\right)
\end{equation}
The corresponding generating vectors for the lattice in
Fig.~\ref{fig:fuerzapos}, denoted as V,  are:
\begin{equation}
{\bf v}_{V,1}=\left(\begin{array}{c} a \\ 0
\end{array}\right) ;\quad {\bf v}_{V,2}=\left(\begin{array}{c} a/2 \\
b/4
\end{array}\right)
\end{equation}

One can show that, for a square array of defects, i.e., $a=b$,
lattice H is a 90$^{\rm o}$ degree rotation of lattice V and the
interaction energy is obviously identical.

The ground state of the system without triangular defects is a
regular triangular lattice T with vortex density $\sigma$, whose
generating vectors can be written as:
\begin{equation}
{\bf v}_{T,1}=\left(\begin{array}{c} \sqrt{\frac{2}{\sigma\sqrt{3}}} \\
0 \end{array}\right)
;\quad {\bf v}_{T,2}=\left(\begin{array}{c} \sqrt{\frac{1}{2\sigma\sqrt{3}}}\\
\sqrt{\frac{\sqrt{3}}{2\sigma}}
\end{array}\right)
\end{equation}

For an aspect ratio $b/a=\sqrt{3}/2$, the vortex density is
$\sigma=8/(a^2\sqrt{3})$ and lattice H coincides with the regular
triangular lattice T. For $b/a=2/\sqrt{3}$, one can also prove that
lattice V becomes regular. Therefore, any deviation (as in our
experiment) from these aspect ratios induces a distortion in the
lattices in order to conform with the periodicity of the array. In
fact, since the aspect ratio of the array in the experiment and simulations
$b/a=746/770=0.969$ is closer to $\sqrt{3}/2$ rather than to
$2/\sqrt{3}$, H is less distorted than V.

Another trivial way of conforming to the periodicity of the array
with $n=4$ vortices per defect, is the rectangular lattice R
generated by:
\begin{equation}
{\bf v}_{R,1}=\left(\begin{array}{c} a/2 \\ 0
\end{array}\right) ;\quad {\bf v}_{R,2}=\left(\begin{array}{c} 0 \\
b/2
\end{array}\right)
\end{equation}

\begin{figure}[tbp]
\center
\includegraphics[width=7cm]{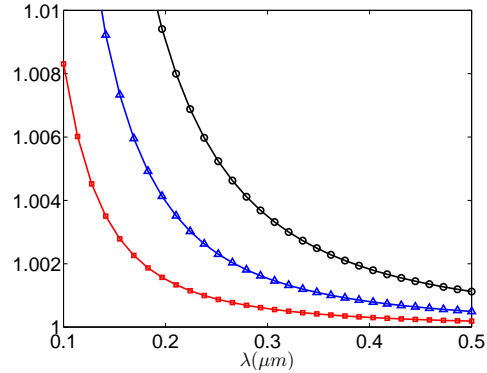}
\caption{Interaction energy for the H, V, and R lattices compared to
the regular triangular one, T: $E_H/E_T$ (red squares), $E_V/E_T$ (blue triangles),
$E_R/E_T$ (black circles). Computed numerically to convergence, including 10000 neighbours.} \label{fig:energias}
\end{figure}

We have evaluated the interaction energy, $E_H,E_V,E_T,E_R$, of
these four lattices for the values $a=770\,{\rm nm}$, $b=746\,{\rm nm}$, and
$\sigma=4/(ab)$ corresponding both to the experiment and simulations.
 The energies are very close to each other. In Fig.
\ref{fig:energias} we plot the energies compared to the ground state
T, i.e., the ratios $E_H/E_T$, $E_V/E_T$, $E_R/E_T$, as a function
of $\lambda$. As expected, the minimal interaction energy
corresponds to the regular lattice T, followed by the less distorted
triangular one, which is H for the aspect ratio 0.969 in the
experiment. Since this aspect ratio is close to 1, the energies of H
and V are very similar. The rectangular lattice, on the other hand,
is the most energetic configuration and, consequently, the most
unstable. Accordingly, we have never seen this type of configuration
in our simulations.

We have of course to consider the pinning energy, which favors, in
absence of external force, lattice H, becoming the ground state of
the system, as observed in the simulations (see
Fig.~\ref{fig:fundamental}). However,  the above calculations point
out that the energy cost, due to vortex interaction,  of rotating
the H configuration to V is relatively small. Since the pinning
energy is, by construction, comparable to the interaction energy, it
is reasonable to explain the response of the system as a competition
between lattices H and V.

It can also be drawn from the previous discussion that a substrate designed with a different geometry (for example, a lattice of defects with an aspect ratio closer to $2/\sqrt{3}$)  could favour the appearance of one of the lattices over the other and this might prevent the transition between them, affecting current reversal. 

\section{Conclusions}

We have shown that  current reversal in superconducting ratchets can
be explained as a dynamical reconfiguration of the vortex lattice.
We have presented numerical evidence of this mechanism and also
showed that the interaction energy of the involved lattices is close
enough to allow such reconfiguration.

The present work prompts further research in several lines. First,
the extension of experimental and theoretical studies on lattice
transitions \cite{PRL99Martin,reconfig1,reconfig2} to the case of
moving lattices. Secondly, the analysis of the interplay between
rectification and lattice configuration. We have shown that the
intensity and direction of rectification can be affected not only by
the shape but also by the array of defects. This could help to
design rectifiers with specific responses to external forces as well
as become a tool to study lattice configurations. For instance,
magnetoresistance, i.e., the response of the system to dc currents,
is commonly used to reveal the underlying vortex lattice in
superconducting films \cite{science,PRL97Martin}. Our work indicates
that the response to ac currents in the presence of asymmetric
artificial defects can also provide valuable information on the
lattice structure.

Finally, similar ratchet mechanisms could be observed in other type
of systems. The main ingredient is a lattice or complex object
undergoing conformational changes as a response to  external forces
and in order to adapt to some anisotropic substrate. This
conformation changes will depend on the sign of the force and can
amplify or modify the rectification properties of the substrate. We
believe that this type of generic ratchet effect could be observed
in systems with interacting particles, like colloidal suspensions or
granular media.

We acknowledge funding support by Spanish Ministerio de
Educaci\'{o}n y Ciencia, grants NAN04-09087, FIS05-07392,
MAT05-23924E, and MOSAICO; CAM grant S-0505/ESP/0337, and
UCM-Santander. Also, computer simulations of this work were performed partly at the ``Cluster de c\'alculo  para T\'ecnicas F\'isicas'' of UCM, funded in
part by UE-FEDER program and in part by UCM and in the ``Aula Sun Microsystems'' at UCM.

\end{document}